\documentclass[preprint,12pt]{elsarticle}

\usepackage{amssymb}
\usepackage{endfloat} 

\usepackage{graphicx}  
\usepackage{epstopdf}
\usepackage{amssymb}
\usepackage[latin1]{inputenc}
\usepackage{epsfig}
\usepackage{amsmath}   

\begin{document}

\begin{frontmatter}

\title{Relating the spectrum of cardiac signals to the spatiotemporal dynamics of cardiac sources}

\author{Jes\'us Requena-Carri\'on\corref{cor1}}
\ead{jesus.requena@urjc.es}
\author{Ferney A.~Beltr\'{a}n-Molina}
\ead{ferney.beltran@urjc.es}
\author{Antonio G.~Marques}
\ead{antonio.garcia.marques@urjc.es}

\cortext[cor1]{Corresponding author\\
			University Rey Juan Carlos, Department of Signal Theory \& Communications\\
			Camino del Molino s/n, Dept III, D207, Fuenlabrada, Madrid, Spain 28943 \\
              		Tel.: +34-91-488-8463, Fax: +34-91-488-7500\\}
\address{University Rey Juan Carlos, Fuenlabrada, Madrid, Spain 28943}

\begin{abstract}
An increasing number of studies use the spectrum of cardiac signals for analyzing the spatiotemporal dynamics of complex cardiac arrhythmias. However, the relationship between the spectrum of cardiac signals and the spatiotemporal dynamics of the underlying cardiac sources remains to date unclear. In this paper, we derive a mathematical expression relating  the spectrum of cardiac signals to the spatiotemporal dynamics of cardiac sources and the measurement characteristics of the lead systems. Then, by using analytical methods and computer simulations we analyze the spectrum of cardiac signals measured by idealized lead systems during correlated and uncorrelated spatiotemporal dynamics. Our results show that lead systems can have distorting effects on the spectral envelope of cardiac signals, which depend on the spatial resolution of the lead systems and on the degree of spatiotemporal correlation of the underlying cardiac sources. In addition to this, our results indicate that the spectral features that do not depend on the spectral envelope, such as the dominant frequency, behave robustly against different choices of lead systems.
\end{abstract}


\end{frontmatter}

\section{\label{sec:introduction} Introduction}

Fibrillation is a complex cardiac arrhythmia whose spatiotemporal characteristics remain poorly understood. Traditionally fibrillation has been described as random and disorganized, since it induces highly irregular traces in the electrocardiogram (ECG) signal. However, with the application of nonlinear dynamics theory to the investigation of cardiac arrhythmias, and the development of optical and electrical mapping techniques, it has been suggested that fibrillation can possess some degree of spatiotemporal regularity \cite{Hoekstra95,Jalife00}. This view has led in a natural way to study fibrillation based on the spectrum of cardiac signals such as the ECG and intracardiac electrograms (EGM). Spectral features of cardiac signals have been proposed as experimental indices for detecting ventricular fibrillation (VF) \cite{Barro89}, for quantifying the degree of spatiotemporal organization of atrial fibrillation (AF) \cite{Everett01a} and for predicting the success of defibrillation shocks \cite{Strohmenger97,Eftestol00,Jekova04}. Also, intracardiac mapping techniques have been combined with dominant frequency (DF) analysis to study the spatiotemporal characteristics of fibrillation. This method, known as DF mapping, has revealed spatiotemporal regularities during AF in both animal models \cite{Skanes98,Mandapati00,Mansour01} and in patients \cite{Sanders05,Atienza06} and it is currently regarded as a potential technique to guide AF ablation therapies \cite{Berenfeld11Book}.

Despite the increasing number of studies that use spectral techniques to analyze fibrillation, the meaning of the spectrum of cardiac signals remains to date elusive. Even though individual spectral features of cardiac signals have been linked to spatiotemporal characteristics of cardiac rhythms \cite{Fischer07,Ng07a,Zlochiver08}, the relationship between the spectrum of cardiac signals and the spatiotemporal characteristics of cardiac rhythms has not been thoroughly investigated.  In addition to this, the effects of lead systems on the spectrum of cardiac signals are not well understood, and consequently it is not clear how the spectra of cardiac signals measured by different lead systems relate to one another. The elucidation of the relationship between the spectra of cardiac signals measured by different lead systems is of technical and clinical interest in the context of fibrillation, since it would contribute to the development of novel, improved methods of DF cardiac mapping, such as non-contact intracardiac electrical mapping \cite{Lin07,Gojraty09} and non-invasive surface ECG mapping \cite{Berenfeld10}.

In this paper, we develop a mathematical formalism for investigating, in a systematic way,  the spectral manifestation of different cardiac rhythms and the spectral effects of the lead systems. By following a multivariate signal analysis approach, we identify the mathematical relationship between the spectrum of cardiac signals and the spatiotemporal dynamics of the underlying cardiac rhythms. Our formalism allows us to derive theoretical results which are relevant for the analysis and interpretation of the spectrum of cardiac signals, and for devising spectral methods for investigating the spatiotemporal dynamics of cardiac rhythms.

The organization of this paper is as follows. In Section \ref{sec:2order_sources} we develop our mathematical formalism and connect the spectrum of cardiac signals to the spatiotemporal dynamics of the underlying cardiac rhythms. Physiologically meaningful cases are studied analytically in Section \ref{sec:analytical}, and in a computer simulation environment in Section \ref{sec:simulations}. Finally, Section \ref{sec:discussion} conveys the conclusions of our investigation and the discussion.

\section{\label{sec:2order_sources} Mathematical formalism} 

In this section we present the mathematical formalism for investigating the spectrum of cardiac signals. Firstly, we introduce the lead-field bioelectric model of cardiac sources and signals. Then, based on multivariate signal analysis, we define the autocorrelation and the spectrum of cardiac sources. Finally, we identify the relationship between: the spectrum of cardiac signals, the spatiotemporal dynamics of cardiac sources and the measurement characteristics of lead systems.

Throughout this paper the following notation is used: $\langle\cdot\rangle_t$ denotes time-average, $\mathcal{F}[\cdot]$ is the Fourier Transform operator, $(\ast)$ denotes convolution and $\delta(\cdot)$ is the Dirac's delta. We use the following vector definitions: $\mathbf{1}=[1, 1, 1]^T$ and $\mathbf{0}=[0, 0, 0]^T$.

\subsection{\label{subsec:bioelectric_model} Bioelectric model}

Cardiac sources are the bioelectric processes generated by the heart during contraction. There exist different, equivalent mathematical paradigms to model the activity of cardiac sources, such as the monopole field and the dipole field \cite{Malmivuo95}. In this study, we model cardiac sources as a time-varying dipole field, i.e. as a spatial distribution of time-varying dipoles $\mathbf{J}(v,t)= [J_x(v,t), J_y(v,t), J_z(v,t)]^T$ on a volume $V$, where $v$ denotes a point located inside $V$ and $t$ denotes the time instant. 

The time-varying activity of cardiac sources can be measured by lead systems, producing cardiac signals. Taking the dipole field as our reference description for cardiac sources, we follow a lead-field approach to model cardiac signals \cite{Malmivuo95}. According to the lead-field theory, the cardiac signal $c(t)$ that is induced at a lead system by a dipole field $\mathbf{J}(v,t)$ can be expressed as
\begin{equation}\label{eq:EqSintesis}
c(t)=\int_V{\mathbf{L}^{T}(v) \mathbf{J}(v,t)}{dv}
\end{equation}
where the vector field $\mathbf{L}(v)= [L_x(v), L_y(v), L_z(v)]^T$ is the measurement sensitivity distribution (MSD) and describes the ability of the lead system to measure cardiac dipoles located at $v\in V$. In words, cardiac signals are a weighted linear combination of the underlying cardiac sources.

\subsection{\label{subsec:aut_spec_sources} Autocorrelation and spectrum of cardiac sources}

The autocorrelation of a cardiac source, $\boldsymbol{\rho}(v,w,\tau)$, $\forall v,w\in V$, is defined as the collection of the cross-correlations between all pairs of dipoles in $V$. Since cardiac dipoles are vectorial entities, the cross-correlation between two dipoles consists of the cross-correlations between all three components of each dipole \cite{Papoulis91}. 
In order to define the autocorrelation of a cardiac source, the average dipole field $\bar{\mathbf{J}}(v)$ needs to be introduced:
\begin{eqnarray}\label{eq:average_dipole}
\bar{\mathbf{J}}(v)&=&\langle\mathbf{J}(v,t)\rangle_t  \nonumber \\
			     &=& [\langle J_x(v,t)\rangle_t, \langle J_y(v,t)\rangle_t, \langle J_z(v,t)\rangle_t]^T. 
\end{eqnarray}
Based on $\bar{\mathbf{J}}(v)$, we define the zero-average dipole field $\mathbf{J'}(v,t)=\mathbf{J}(v,t)-\bar{\mathbf{J}}(v)$, so that $\langle\mathbf{J'}(v,t)\rangle_t=\mathbf{0}$. The cross-correlation matrix between two cardiac dipoles $\mathbf{J}(v,t)$ and $\mathbf{J}(w,t)$, where $v,w\in V$, is then defined as
\begin{eqnarray}\label{eq:autocorrelation_source}
&&\boldsymbol{\rho}(v,w,\tau)=\langle\mathbf{J'}(v,t+\tau)\mathbf{J'}^{T}(w,t)\rangle_t \nonumber \\
&&= \left( \begin{array}{ccc}
{\rho}_{xx}(v,w,\tau) & {\rho}_{xy}(v,w,\tau) & {\rho}_{xz}(v,w,\tau) \\
{\rho}_{yx}(v,w,\tau) & {\rho}_{yy}(v,w,\tau) & {\rho}_{yz}(v,w,\tau) \\
{\rho}_{zx}(v,w,\tau) & {\rho}_{zy}(v,w,\tau) & {\rho}_{zz}(v,w,\tau) 
\end{array} \right).
\end{eqnarray}
Therefore, each entry of $\boldsymbol{\rho}(v,w,\tau)$ contains the cross-correlation between one component of $\mathbf{J}(v,t)$ and one component of $\mathbf{J}(w,t)$. For instance, matrix entry ${\rho}_{zy}(v,w,\tau)$ is $\langle J'_z(v,t+\tau) J'_y(w,t)\rangle_t$. Also, the average power of dipole component  $J'_x(v,t)$ is by definition $P_x(v)={\rho}_{xx}(v,v,0)$, and analogous expressions can be obtained for the average power of dipole components $J'_y(v,t)$ and $J'_z(v,t)$.

The spectrum of a cardiac source, $\boldsymbol{\sigma}(v,w,f)$, $\forall v,w\in V$, corresponds to the collection of the cross-spectra between all pairs of dipoles in $V$, and is defined as
\begin{eqnarray}\label{eq:spectrum_source}
&&\boldsymbol{\sigma}(v,w,f) =\mathcal{F}[\boldsymbol{\rho}(v,w,\tau)] \nonumber \\
&&=\left( \begin{array}{ccc}
{\sigma}_{xx}(v,w, f) & {\sigma}_{xy}(v,w, f) & {\sigma}_{xz}(v,w, f) \\
{\sigma}_{yx}(v,w, f) & {\sigma}_{yy}(v,w, f) & {\sigma}_{yz}(v,w, f) \\
{\sigma}_{zx}(v,w, f) & {\sigma}_{zy}(v,w, f) & {\sigma}_{zz}(v,w, f)
\end{array} \right)
\end{eqnarray}
where the operator $\mathcal{F}[\cdot]$ is applied to $\boldsymbol{\rho}(v,w,\tau)$ on a component-by-component basis. For instance, ${\sigma}_{zy}(v,w, f)$ is $\mathcal{F}[{\rho}_{zy}(v,w,\tau)]$. 

We also define the \emph{total cross-correlation} $R_{J}(v,w,\tau)$ between two cardiac dipoles $\mathbf{J}(v,t)$ and $\mathbf{J}(w,t)$ as the sum of the entries of $\boldsymbol{\rho}(v,w,\tau)$ and the \emph{total cross-spectrum} $S_{J}(v,w,f)$ as the Fourier Transform of $R_{J}(v,w,\tau)$. Mathematically, they can be expressed as
\begin{eqnarray}
R_{J}(v,w,\tau) &=&  \mathbf{1}^{T} \boldsymbol{\rho}(v,w,\tau)  \mathbf{1},\label{total_corr}\\
S_{J}(v,w,f)  &=&  \mathbf{1}^{T} \boldsymbol{\sigma}(v,w, f)  \mathbf{1}. \label{total_spectrum}
\end{eqnarray}

Finally, we define the \emph{normalized} cross-correlation $\boldsymbol{\hat{\rho}}(v,w,\tau)$ as the matrix of entries
\begin{equation}\label{eq:normalized_autocorrelation_source}
\hat{\rho}_{ab}(v,w,\tau) = \frac{{\rho}_{ab}(v,w,\tau)}{\sqrt{{\rho}_{aa}(v,v,0){\rho}_{bb}(w,w,0)}}
\end{equation}
where $a,b\in \{x,y,z\}$, and the \emph{normalized} total cross-correlation $\hat{R}_{J}(v,w,\tau)$ as
\begin{equation}\label{normalized_total_corr}
\hat{R}_{J}(v,w,\tau)= \frac{R_{J}(v,w,\tau)}{{\underset{\tau}{\max}}\{R_{J}(v,w,\tau)\}}.
\end{equation}

\subsection{\label{sec:2order_signals} Autocorrelation and spectrum of cardiac signals}

Let $c(t)$ be a cardiac signal measured by applying $\mathbf{L}(v)$ to a cardiac source of autocorrelation $\boldsymbol{\rho}(v,w,\tau)$ and spectrum $\boldsymbol{\sigma}(v,w,f)$ in $V$. Define $c'(t)$ as the cardiac signal $c(t)$ minus its time-average value $\bar{c}=\langle c(t) \rangle_t$,
\begin{equation}\label{c_minus_average}
c'(t)=c(t)-\bar{c}.
\end{equation}
The autocorrelation function $R_{c}(\tau)$ of the cardiac signal $c(t)$ is defined as the following average \cite{Papoulis91}:
\begin{equation}\label{signal_autocorr}
R_{c}(\tau)  = \langle c'(t+\tau)c'(t) \rangle_t,
\end{equation}
and its power spectrum $S_{c}(f)$ is defined as the Fourier Transform of its autocorrelation function,
\begin{equation}\label{signal_spectrum}
S_{c}(f)  = \mathcal{F}[R_{c}(\tau)].
\end{equation}
Based on \eqref{eq:EqSintesis}, the following relationships can be derived between $R_{c}(\tau)$ and $\boldsymbol{\rho}(v,w,\tau)$, and betwen $S_{c}(f)$ and $\boldsymbol{\sigma}(v,w,f)$ (see Appendix A):
\begin{eqnarray}
R_{c}(\tau)&=& \int_{V\times V}{\mathbf{L}^{T}(v) \boldsymbol{\rho}(v,w,\tau) \mathbf{L}(w)}{dvdw},\label{eq:signal_model_autocorr}\\
S_{c}(f) &=&\int_{V\times V}{\mathbf{L}^{T}(v) \boldsymbol{\sigma}(v,w,f) \mathbf{L}(w)}{dv dw}. \label{eq:signal_model_spectrum}
\end{eqnarray}
Equations \eqref{eq:signal_model_autocorr} and \eqref{eq:signal_model_spectrum} reflect the linear relationship between cardiac signals and sources [c.f. \eqref{eq:EqSintesis}], and can be used to gain insight into the nature of the autocorrelation and spectrum of cardiac signals. According to \eqref{eq:signal_model_spectrum}, two factors determine the spectrum of cardiac signals. The first factor is the spectrum $\boldsymbol{\sigma}(v,w,f)$ of cardiac sources. It is worth noting that this is the solely feature of the spatiotemporal dynamics of cardiac sources that manifests on the spectrum of cardiac signals. The second factor is the MSD of the lead system, $\mathbf{L}(v)$. Since $\mathbf{L}(v)$ is specific for each lead system, \eqref{eq:signal_model_spectrum} reveals that cardiac signals measured by different lead systems will in general have different spectra for the same underlying spatiotemporal dynamics.

\section{\label{sec:analytical} Analytical study}

In this section we study analytically the spectral manifestation of two second-order models of cardiac sources, namely the fully correlated (FC) source and the fully uncorrelated (FU) source.  The FC and FU models are physiologically meaningful and can be used to describe the dynamics of, respectively, highly organized and highly disorganized cardiac rhythms. 

We firstly define the autocorrelation and the spectrum of the following models of spatiotemporal dynamics: identically distributed (ID), FC and FU. The ID model is introduced for facilitating the comparison of the spectrum of cardiac signals measured during FC and FU dynamics. Secondly, we define a simple, idealized model of MSD, namely the pulse model. Because of its simplicity, the pulse model is used in the analytical derivations and in the simulation experiments throughout this study. Finally, we derive analytically the spectrum of cardiac signals measured by pulse MSD during FU and FC dynamics.

\subsection{Models of spatiotemporal dynamics}\label{source_models}

We subsequently present the three second-order models of spatiotemporal dynamics that we use in this study, namely the ID, the FC and the FU models.

\subsubsection{Identically distributed spatiotemporal dynamics}
In this model of spatiotemporal dynamics all cardiac dipoles have the same autocorrelation and spectrum, 
\begin{eqnarray}
\boldsymbol{\rho}(v,v,\tau)=\boldsymbol{\rho}(\tau), \label{corr_IDS}\\
\boldsymbol{\sigma}(v,v,f)=\boldsymbol{\sigma}(f). \label{spectrum_IDS}
\end{eqnarray}
By substituting \eqref{corr_IDS} and \eqref{spectrum_IDS} respectively into \eqref{total_corr} and \eqref{total_spectrum}, it can be proved that the total autocorrelation and total spectrum of all the dipoles are also identical, 
\begin{eqnarray}
R_{J}(v,v,\tau)&=&R_{J}(\tau)=\mathbf{1}^{T} \boldsymbol{\rho}(\tau)  \mathbf{1},\label{total_corr_IDS}\\ 
S_{J}(v,v,f)&=&S_{J}(f)=\mathbf{1}^{T} \boldsymbol{\sigma}(f)  \mathbf{1}. \label{total_spectrum_IDS}
\end{eqnarray}
Note that this model only describes the activity of cardiac dipoles individually, and does not specify $\boldsymbol{\rho}(v,w,\tau)$ nor $\boldsymbol{\sigma}(v,w,f)$ for $v\neq w$.

\subsubsection{Fully correlated spatiotemporal dynamics}
This model of spatiotemporal dynamics corresponds to highly regular rhythms, such as sinus rhythm, in which the activity of one dipole $\mathbf{J}(w,t)$ can be expressed as a delayed version of the activity of another dipole $\mathbf{J}(v,t)$,
\begin{equation}\label{def:FCS}
\mathbf{J}(w,t)=\mathbf{J}(v,t-\zeta (v,w)),
\end{equation}
where $\zeta (v,w)$ is defined as the time delay between the activities of dipoles $\mathbf{J}(v,t)$ and $\mathbf{J}(w,t)$. Based on \eqref{def:FCS}, it can be proved (see Appendix B) that FC dynamics are also ID, $\boldsymbol{\rho}(v,v,\tau)=\boldsymbol{\rho}(\tau)$ and $\boldsymbol{\sigma}(v,v,f)=\boldsymbol{\sigma}(f)$ [cf. \eqref{corr_IDS} and \eqref{spectrum_IDS}],  and that the autocorrelation and the spectrum of FC sources can be expressed as:
\begin{eqnarray}
\boldsymbol{\rho}(v,w,\tau) &=&\boldsymbol{\rho}(\tau - \zeta (v,w)), \label{eq:autocorrelation_correlated1}\\
\boldsymbol{\sigma}(v,w,f) &=& \boldsymbol{\sigma}(f) \exp[-j2\pi f \zeta (v,w)]. \label{eq:spectrum_correlated}
\end{eqnarray}
Consequently, FC sources are completely characterized by $\boldsymbol{\rho}(\tau)$, $\boldsymbol{\sigma}(f)$ and $\zeta (v,w)$.

\subsubsection{Fully uncorrelated spatiotemporal dynamics}
This model of spatiotemporal dynamics constitutes an idealization of highly irregular and disorganized rhythms, such as fibrillation, in which there is no second-order relationship between the temporal activity of any pair of cardiac dipoles. The autocorrelation and the spectrum of a FU source are defined as
\begin{eqnarray}
\boldsymbol{\rho}(v,w,\tau)=\boldsymbol{\rho}(v,v,\tau)\delta(v-w),\label{eq:autocorrelation_uncorrelated} \\
\boldsymbol{\sigma}(v,w,\tau)=\boldsymbol{\sigma}(v,v,\tau)\delta(v-w).\label{eq:spectrum_uncorrelated}
\end{eqnarray}
In words, the cross-correlation between two cardiac dipoles $\mathbf{J}(w,t)$ and $\mathbf{J}(v,t)$, where $v\neq w$, is null.

\subsection{Pulse model of MSD}

In this section we define one simple, idealized model of MSD, namely the pulse model. The pulse model describes a lead system that measures with the same sensitivity every dipole within a region $V_0$ of the volume source $V$, while rejecting the rest. Mathematically, this model is defined as
\begin{equation}
\label{def:pulso}
\mathbf{L}_{V_0}(v)= 
\begin{cases}
    \begin{tabular}{ll}
    $\mathbf{1}$ & if $v\in V_0$\\
    $\mathbf{0}$ & otherwise
        \end{tabular}
 \end{cases}.
\end{equation}
The pulse model can be treated as an approximation of physical MSD that effectively concentrate their measurement in a region $V_0$. 

For the subsequent analysis it is also convenient to quantify the spatial resolution (SR) of pulse leads. The SR can be defined as the region of the cardiac source that contributes the most to the measured signal. In this study, we quantify the SR of pulse leads by introducing the notion of the lead equivalent volume (LEV), which is defined as the relative size of $V_{0}$ to the size of $V$,
\begin{equation}
LEV=\frac{\int_{V_{0}}{dv}}{\int_{V}{dv}}=\frac{M_{V_{0}}}{M_V},\label{eq:SR}
\end{equation}
where $M_{V_{0}}$ and $M_V$ are the sizes of $V_{0}$ and $V$ respectively. Thus, for local measurements the LEV is close to zero, whereas for global measurements where $V_0\simeq V$, the LEV is close to one.

\subsection{Spectrum of cardiac signals during FC and FU dynamics}

\subsubsection{Fully correlated spatiotemporal dynamics}
We subsequently derive the spectrum of cardiac signals measured by pulse leads during FC dynamics.  By substituting \eqref{def:pulso} and \eqref{eq:spectrum_correlated} into \eqref{eq:signal_model_spectrum} and using \eqref{total_spectrum_IDS}, we can express the spectrum $S_{c}(f)$ in terms of the total spectrum $S_{J}(f)$ and the time-delay $\zeta (v,w)$:
\begin{eqnarray}\label{spectrum_FC_CS_1}
S_{c}(f) 		&=& \int_{V\times V}{\mathbf{L}_{V_0}^T(v)  \boldsymbol{\sigma}(v,w,f) \mathbf{L}_{V_0}(w)}{dvdw} \nonumber \\
			 &=& \int_{V_0\times V_0}{ \mathbf{1}^{T} \boldsymbol{\sigma}(f) \exp[-j2\pi f \zeta (v,w)]  \mathbf{1} }{dvdw} \nonumber \\
			 &=& \mathbf{1}^{T} \boldsymbol{\sigma}(f) \mathbf{1} \int_{V_0\times V_0}{  \exp[-j2\pi f \zeta (v,w)] }{dvdw} \nonumber \\
			 &=& S_{J}(f)  \int_{V_0\times V_0}{  \exp[-j2\pi f \zeta (v,w)] }{dvdw}. 
\end{eqnarray}
We can integrate \eqref{spectrum_FC_CS_1} with respect to $\zeta$ by introducing the time-delay density function (TDDF) over $V_0$, $F_{V_0}(\zeta)$. The TDDF $F_{V_0}(\zeta)$ describes the frequency that a time-delay $\zeta$ is observed between two dipoles in $V_0$, and can be calculated as follows. Consider all the pairs of dipoles that reside in $V_0$. Given a time delay $\zeta_1$, select all the pairs of dipoles whose activities are delayed by $\zeta_1$. Then,  the quantity $F_{V_0}(\zeta_1)$ corresponds to the density (\emph{fraction}) of those pairs of dipoles, with respect to all the pairs of dipoles in $V_0$. Based on $F_{V_0}(\zeta)$, we can express \eqref{spectrum_FC_CS_1} as
\begin{eqnarray}\label{spectrum_FC_CS_2}
S_{c}(f) &=& S_{J}(f)  \int_{-\infty}^{\infty}{  \exp[-j2\pi f \zeta] F_{V_0}(\zeta)}{d\zeta}  \nonumber \\
  			 &=& S_{J}(f)  \mathcal{F}[ F_{V_0}(\zeta) ] \nonumber \\
			 &=& S_{J}(f)  S_{V_0}(f).
\end{eqnarray}
In general, time delays $\zeta$  are expected to be short for small $V_0$, i.e. small LEV [c.f. \eqref{eq:SR}]. Hence, the smaller $V_0$, the more concentrated $F_{V_0}(\zeta)$ around  $\zeta=0$ and the broader $S_{V_0}(f)$. Therefore \eqref{spectrum_FC_CS_2} reveals that cardiac signals from local measurements (small LEV) during FC dynamics have broad bandwidths, whereas cardiac signals from global measurements (large LEV) have narrow bandwidths. Equation \eqref{spectrum_FC_CS_2} is a natural result, since the integral in \eqref{eq:EqSintesis} averages \emph{delayed} signals at different points within the effective measurement region $V_0$. As a result, when the LEV is large the lead system acts as a low-pass filter. By contrast, when $V_0$ is small little averaging is performed and the distortion introduced by the lead system is low.

\subsubsection{Fully uncorrelated spatiotemporal dynamics}

We now derive the spectrum of cardiac signals measured by pulse leads during FU dynamics. By substituting \eqref{eq:spectrum_uncorrelated} and  \eqref{def:pulso} into \eqref{eq:signal_model_spectrum}, and using \eqref{total_spectrum}, the spectrum can be written as
\begin{eqnarray}\label{eq:spectrum_FUS}
S_{c}(f) 		&=& \int_{V\times V}{\mathbf{L}_{V_0}^{T}(v) \boldsymbol{\sigma}(v,w,f) \mathbf{L}_{V_0}(w)}{dvdw} \nonumber \\
			&=& \int_{V_0}{ \mathbf{1}^{T} \boldsymbol{\sigma}(v,v,f) \mathbf{1}}{dv} \nonumber \\
			&=& \int_{V_0}{ S_{J}(v,v,f)}{dv}.
\end{eqnarray}
In words, the spectrum of cardiac signals measured during FU dynamics is an average of the total spectra of the dipoles contained within $V_0$. If the FU dynamics is also assumed to be ID (ID-FU), then $S_{J}(v,v,f)=S_{J}(f)$ [cf. \eqref{total_spectrum_IDS}] and \eqref{eq:spectrum_FUS} reduces to
\begin{equation}\label{eq:spectrum_FUS_IDS}
S_{c}(f) 	=  \int_{V_0}{S_{J}(f)}{dv}  = S_{J}(f) M_{V_0}
\end{equation}
Hence, except for the factor of $M_{V_0}$, the spectra of cardiac signals measured by pulse leads during ID-FU dynamics are identical to the total spectrum $S_{J}(f)$.

\section{\label{sec:simulations} Simulation experiments}

In this section we present the results obtained in a computer simulation environment, designed for investigating the spectral manifestation of cardiac rhythms and the spectral effects of lead systems. Firstly, we describe the following aspects of our simulation experiments: the cardiac source model, how we simulated FC and FU dynamics, and how we synthesized cardiac signals. Secondly, we use multivariate signal analysis to describe the simulated FC and FU dynamics. Thirdly, we analyze the TDDF of the simulated FC dynamics and its relationship with the spectrum of measured cardiac signals. Finally, we analyze the effects of different pulse leads on the spectrum of cardiac signals during FC and FU simulated dynamics.

\subsection{Simulation set-up}

Cardiac activity was modeled by following a probabilistic cellular automata approach proposed in \cite{Alonso-Atienza05} which allows simulating complex spatiotemporal dynamics such as fibrillatory conduction and rotors. In this model, cardiac tissue is described as a grid of connected, excitable cells, characterized by a time-varying transmembrane voltage. Cell excitation produces a transient change in the transmembrane voltage, causing excitation in the neighboring cells. This activity results in excitation waves traveling through the cardiac tissue.

\begin{figure*} [t]
\raggedright
\begin{tabular}{cc}
      \begin{tabular}{c} 
      \epsfig{file = 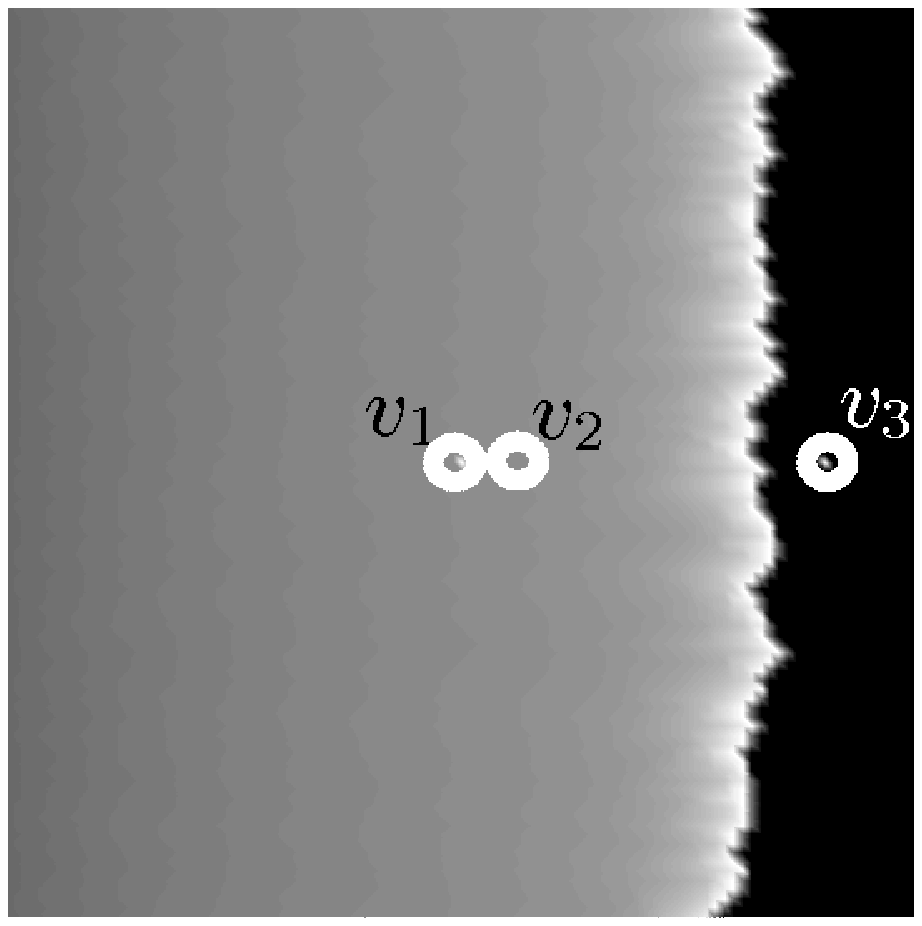,width = 5cm} \\(a)\\ \\
      \epsfig{file = 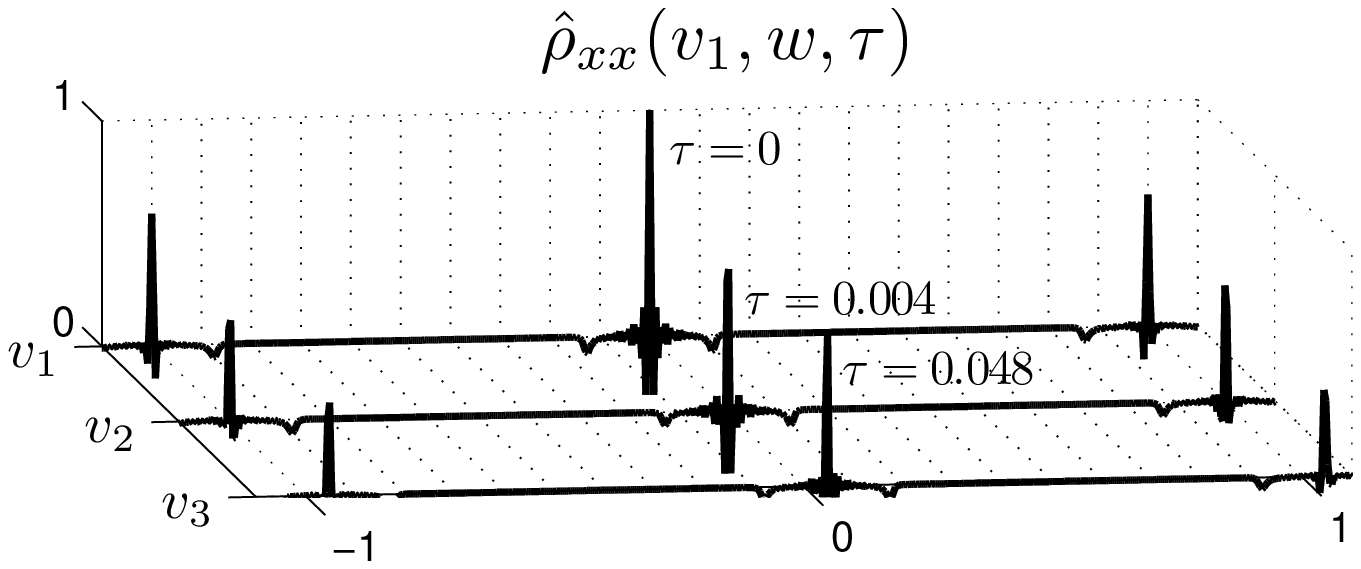,width = 5cm}  \\
      \epsfig{file = 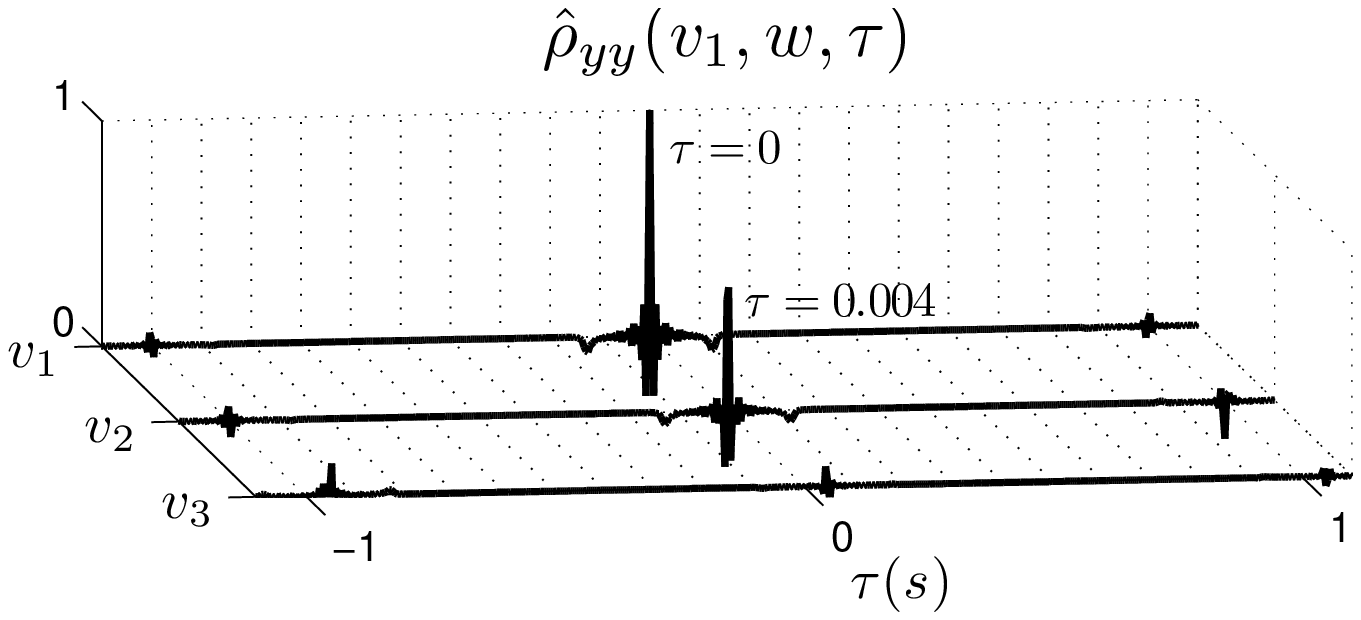,width = 5cm}  \\(c)\\ \\
      \epsfig{file = 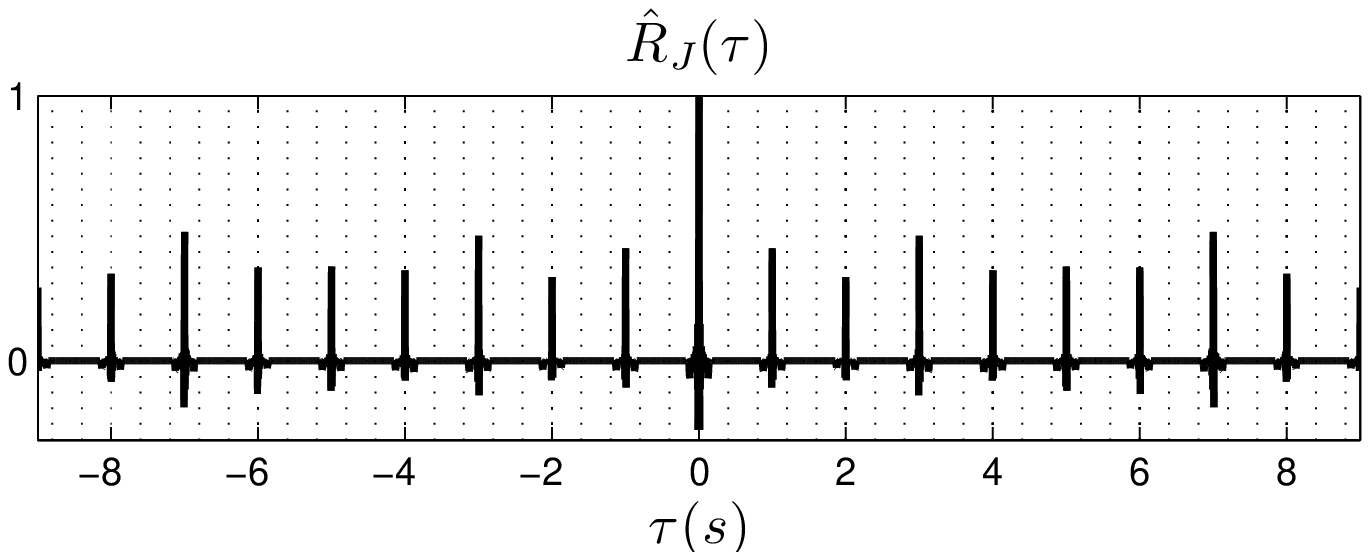,width = 5cm}  \\(e)
      \end{tabular}&
      \begin{tabular}{c} 
      \epsfig{file = 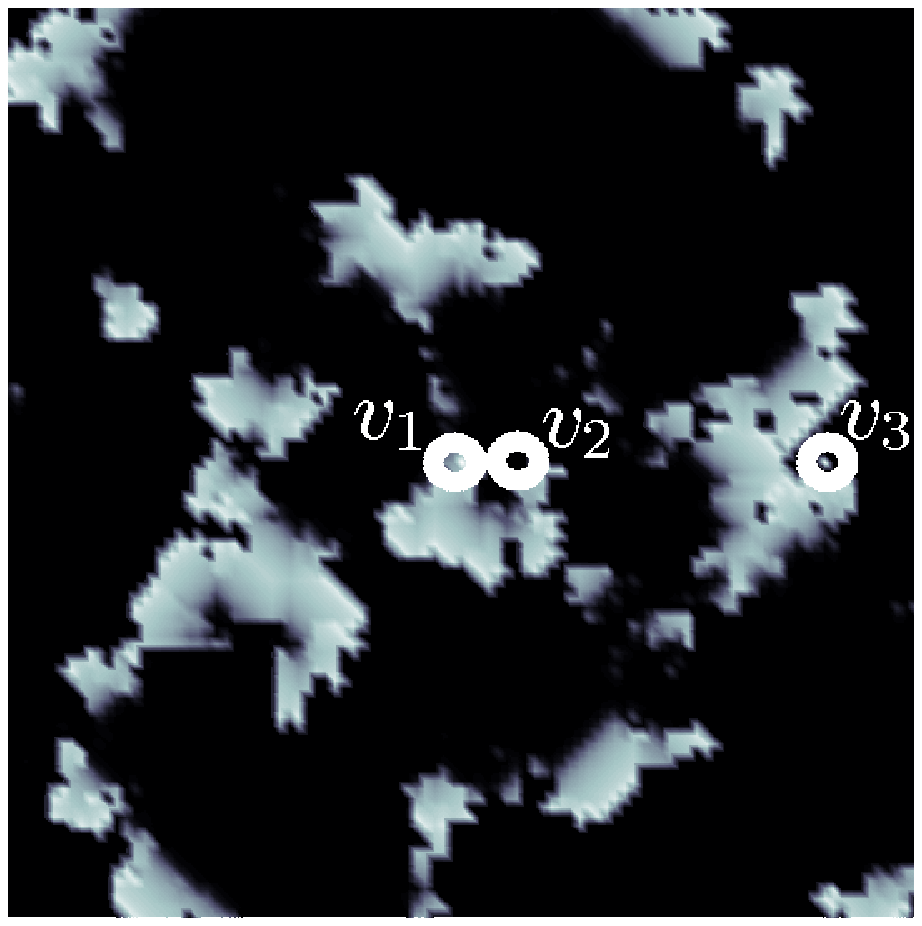,width = 5cm} \\(b)\\ \\
      \epsfig{file = 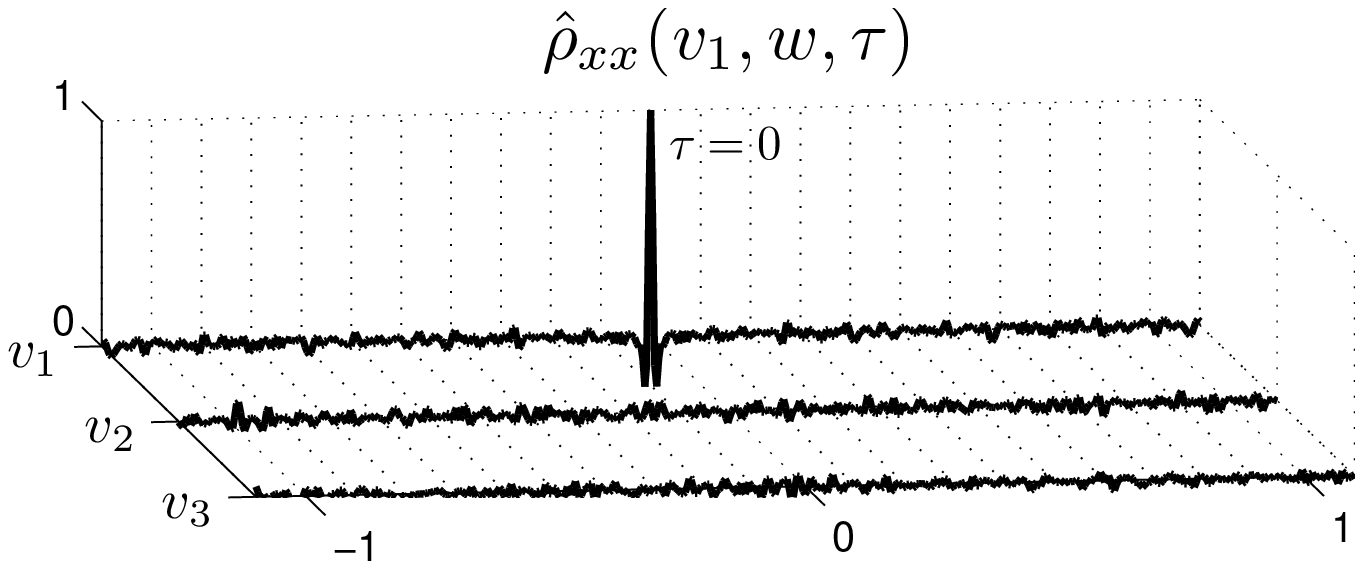,width = 5cm}  \\
      \epsfig{file = 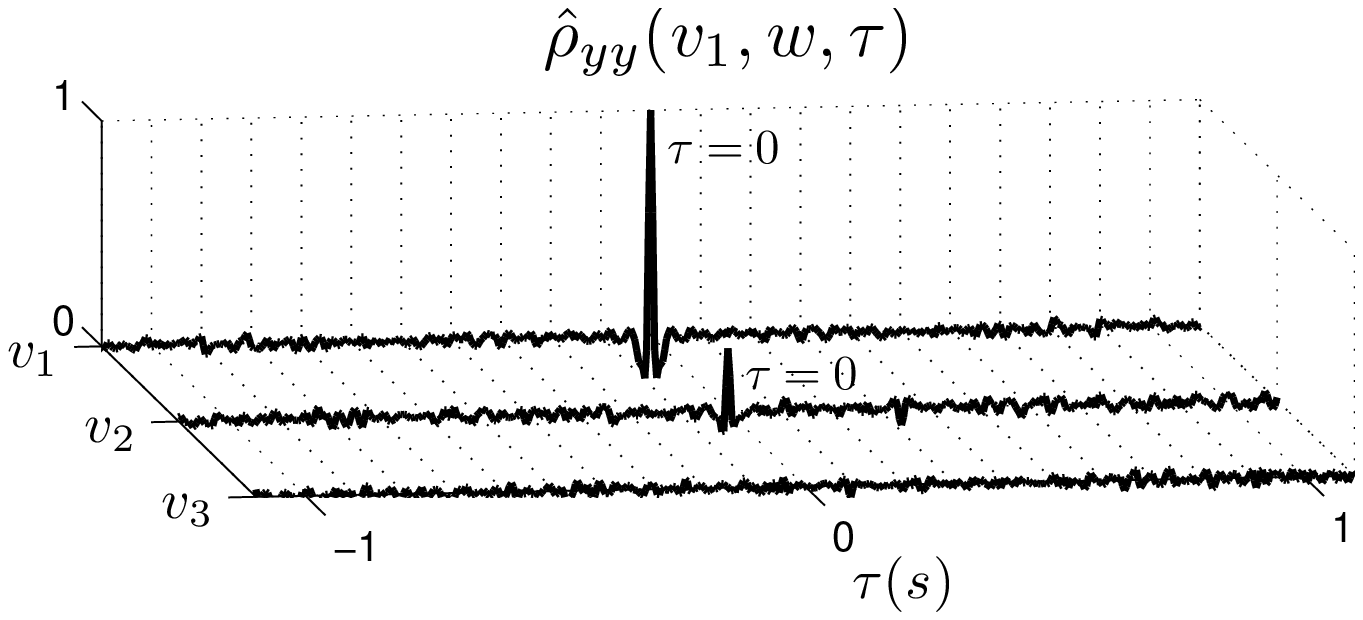,width = 5cm}  \\(d)\\ \\
      \epsfig{file = 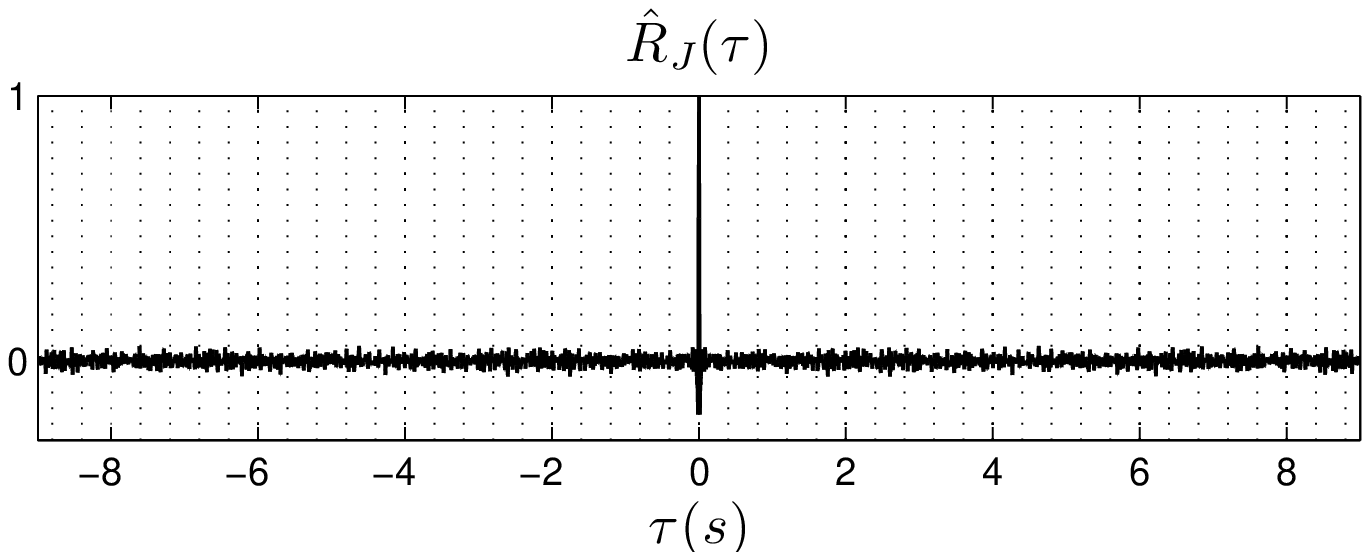,width = 5cm}  \\(f)
      \end{tabular}
\end{tabular}
\caption{Analysis of simulated fully correlated (FC) and fully uncorrelated (FU) dynamics. (See text for explanation.) Voltage maps during (a) FC and (b) FU dynamics, showing source locations $v_1$, $v_2$ and $v_3$. Normalized correlations ${\hat{\rho}}_{xx}(v_1,w,\tau)$ and ${\hat{\rho}}_{yy}(v_1,w,\tau)$, where $w=v_1,v_2,v_3$, during (c) FC and (d) FU dynamics. Normalized total autocorrelation $\hat{R}_{J}(\tau)$ during (e) FC and (f) FU dynamics.
}
\label{fig:Plano_rho}
\end{figure*}

Based on this model, a 2D square sample of cardiac tissue consisting of $101\times 101$ cells was defined and two types of spatiotemporal dynamics were simulated. The first type of dynamics was generated by stimulating one side of the tissue sample at a rate of 1 Hz. This resulted in a FC dynamics consisting of a regular train of plane excitation waves traveling along the $x$ axis (Fig. \ref{fig:Plano_rho} (a)). The second type of dynamics corresponded to a chaotic excitation pattern characteristic of FU sources. This spatiotemporal dynamics consisted of multiple, fragmented excitation waves, traveling randomly in every direction (Fig. \ref{fig:Plano_rho} (b)).

By computing the voltage difference between neighboring cells, a 2D time-varying distribution of cardiac dipoles $\mathbf{J}(v,t)= [J_x(v,t), J_y(v,t)]^T$ was simulated. The amplitude of the simulated cardiac dipoles was highest at the wavefronts, where the voltage gradient is largest, and their direction was as expected aligned with the direction of propagation of excitation waves. In order to synthesize cardiac signals, eight square pulse MSD $A_n$, $1\leq n\leq 8$, of increasing size and centered at the middle of the tissue sample were implemented. Table \ref{Tabla_Pulse} shows the sizes of each pulse MSD along with their LEV, and Fig. \ref{fig:Plano_histo} (b) depicts for illustrative purposes pulses $A_5$ and $A_8$. Finally, a set of cardiac signals $c_n(t)$ were synthesized by applying each pulse $A_n$ to the simulated time-varying dipole distributions. 

\begin{table}[b]
\caption{Definition and lead equivalent volume (LEV) of the measurement sensitivity distributions used in the simulations and bandwidth of the corresponding simulated signals during fully correlated (FC) and fully uncorrelated (FU) dynamics}
    \begin{tabular}{ccccc}
	\hline\noalign{\smallskip}
	Lead& Dimensions&LEV (\%)&$BW_{FC}$ (Hz)&$BW_{FU}$ (Hz)\\
	\noalign{\smallskip}\hline\noalign{\smallskip}
	$A_1$&$5\times5$& 0.2&114&79\\
	$A_2$&$9\times9$& 0.8&77&63\\
	$A_3$&$13\times13$& 1.6&85&61\\
	$A_4$&$17\times17$& 2.8&70&60\\
	$A_5$&$21\times21$& 4.3&60&58\\
	$A_6$&$41\times41$& 16.4&35&54\\
	$A_7$&$61\times61$& 36.4&28&54\\
	$A_8$&$81\times81$& 64.3&24&54\\\noalign{\smallskip}
	\hline
    \end{tabular}
    \label{Tabla_Pulse}
\end{table}

\subsection{Analysis of the spatiotemporal dynamics}

In order to analyze the spatiotemporal characteristics of both simulated dynamics, we obtained the average power of each dipole component, together with the normalized cross-correlation [c.f \eqref{eq:normalized_autocorrelation_source}] and the normalized total cross-correlation [c.f \eqref{normalized_total_corr}] between every pair of dipoles.

\subsubsection{Fully correlated spatiotemporal dynamics}
In this dynamics, the normalized autocorrelation and total autocorrelation were found to be approximately the same for every cardiac dipole, $\boldsymbol{\hat{\rho}}(v,v,\tau)\simeq\boldsymbol{\hat{\rho}}(\tau)$ and $\hat{R}_{J}(v,v,\tau)\simeq\hat{R}_{J}(\tau)$. Accordingly, the dynamics were classified as ID  [c.f. \eqref{corr_IDS} and \eqref{total_corr_IDS}]. Also, the average power $P_x(v)$ was found to be twice as high as $P_y(v)$. This observation is in agreement with the fact that most of the dipoles were aligned with the direction of propagation of the excitation waves, namely the $x$ axis.

\begin{figure*} [t]
\centering
\begin{tabular}{c}
  \epsfig{file = 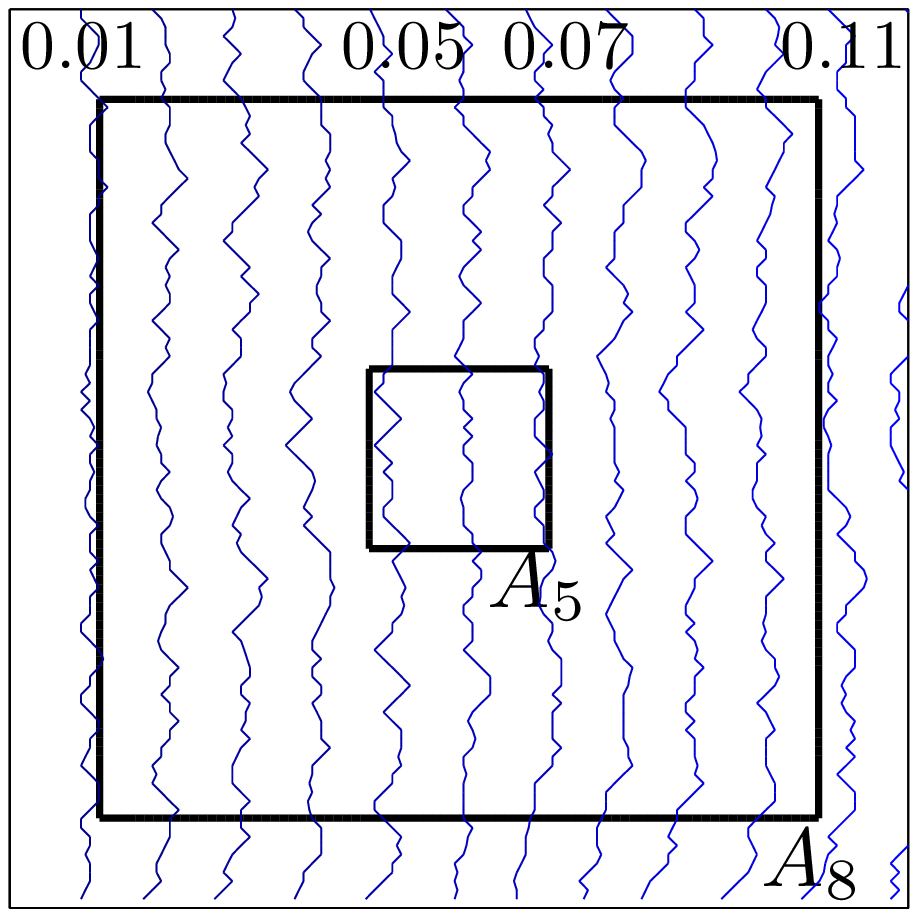,width = 4cm}  \\(a)\\\\
  \epsfig{file = 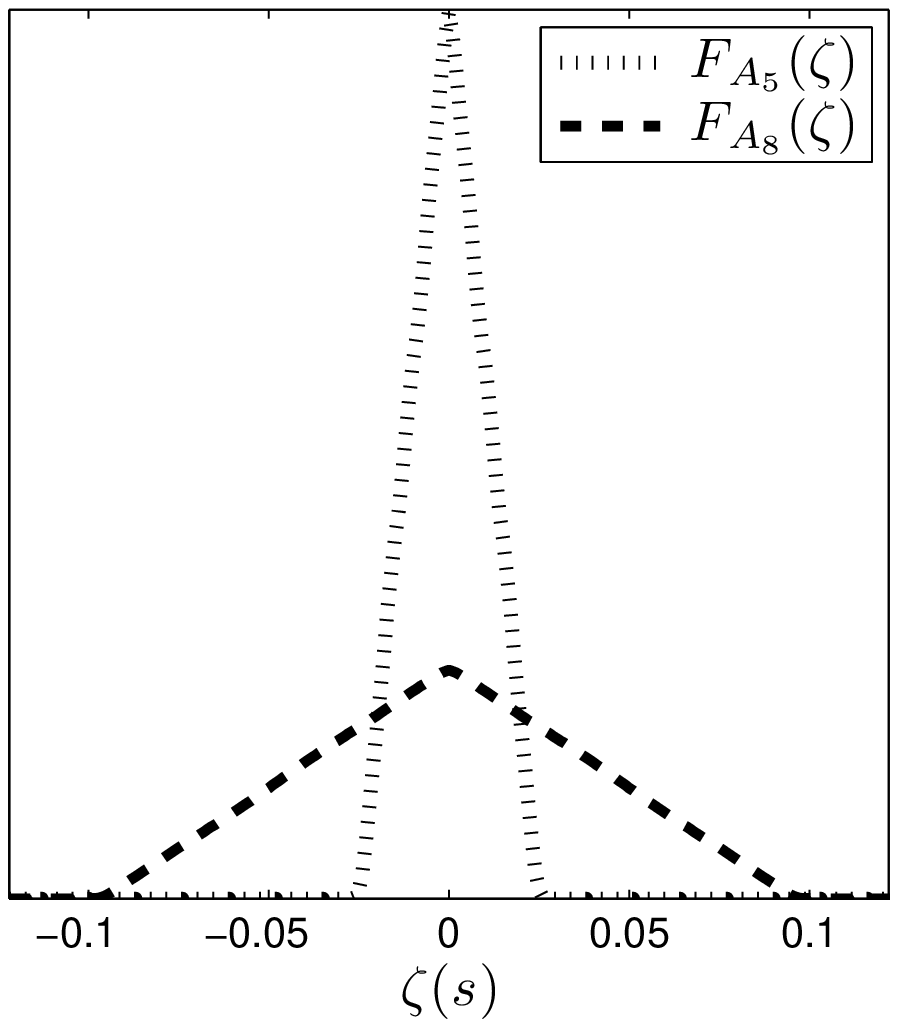,width = 4cm}\\ (b)\\
\end{tabular}
\begin{tabular}{c}
 \epsfig{file = 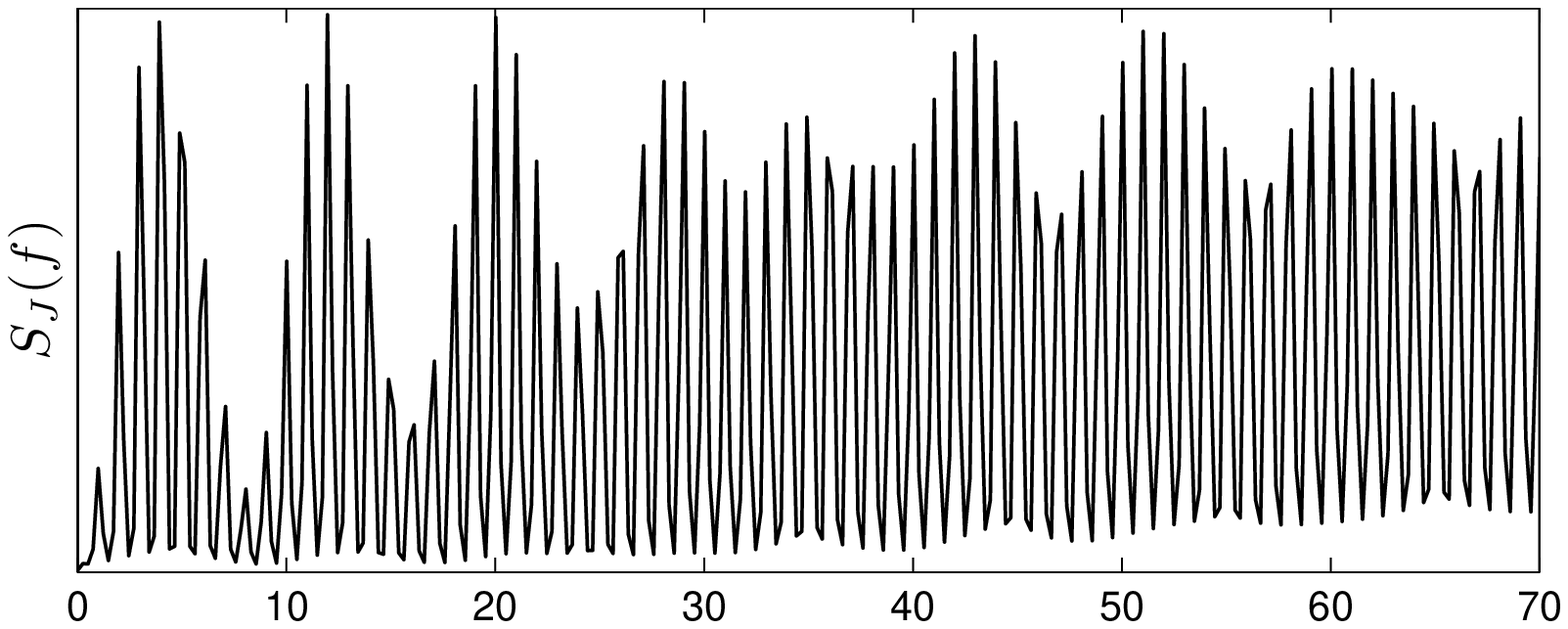,width = 7.2cm} \\
 \epsfig{file = 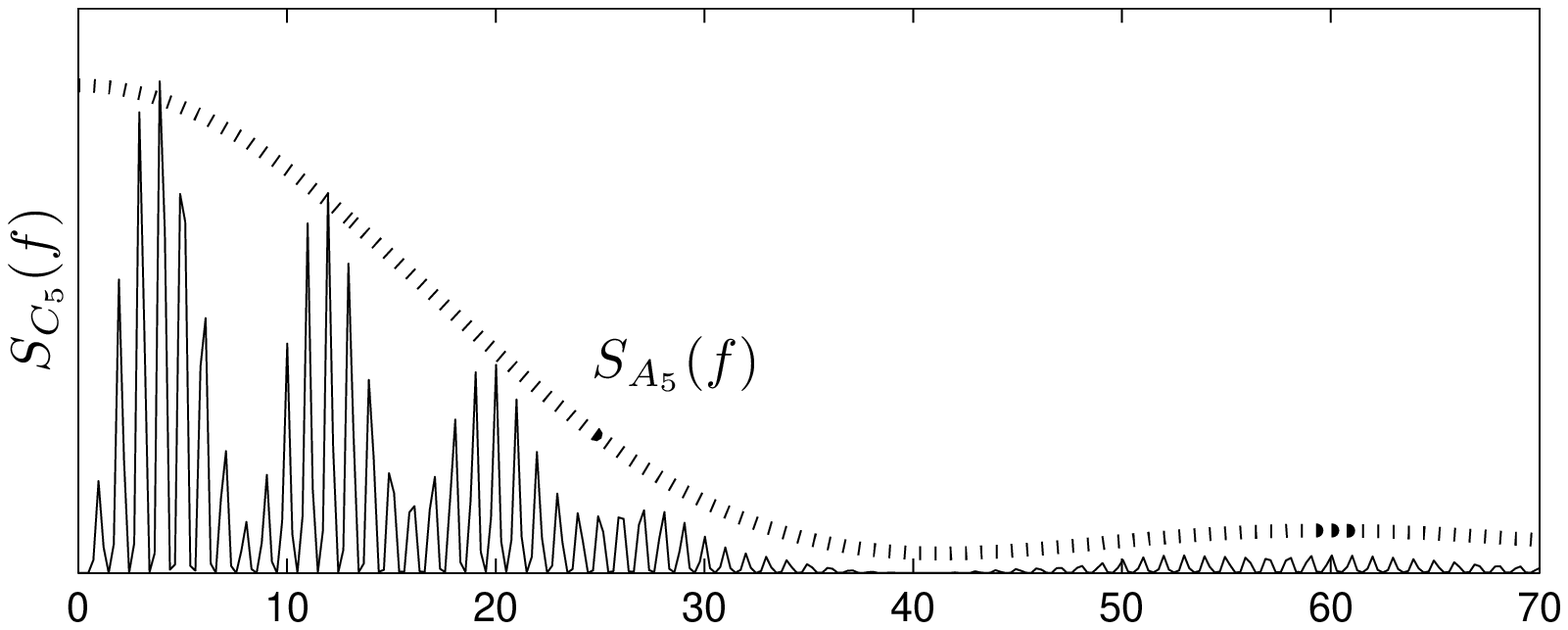,width = 7.2cm}  \\
 \epsfig{file = 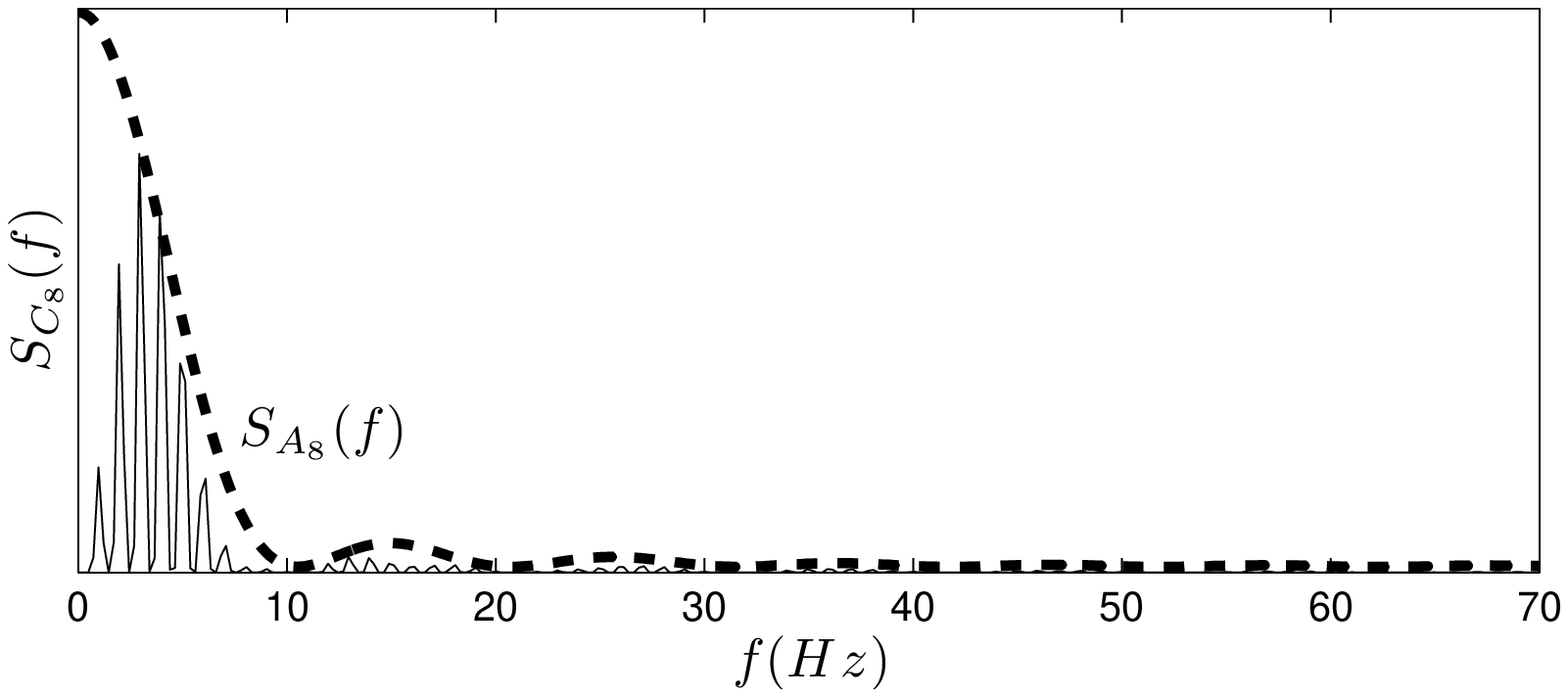,width = 7.2cm}  \\(c)\\
\end{tabular}
\caption{The envelope of cardiac signals measured during fully correlated (FC) dynamics conforms to the Fourier Transform of the time-delay density function (TDDF). (a) An isochronal map of cell excitations during simulated FC dynamics and the regions covered by pulses $A_5$ and $A_8$ are shown. (b) Based on the time delay between cell excitations, the TDDF  corresponding to pulses from leads $A_5$ and $A_8$, $F_{A_5}(\zeta)$ and $F_{A_8}(\zeta)$, are generated. (c) As predicted by \eqref{spectrum_FC_CS_2}, the multiplication of the total spectrum $S_J(f)$ by the Fourier Transforms of the TDDF $S_{A_5}(f)$ and $S_{A_8}(f)$, matches the spectra of signals generated by pulses $A_5$ and $A_8$,  $S_{c_5}(f)$ and $S_{c_8}(f)$.}
\label{fig:Plano_histo}
\end{figure*}

Figure \ref{fig:Plano_rho} (c) shows the normalized cross-correlations between the $x$ and $y$ components of dipole $v_1$, and dipoles $v_1$ (itself), $v_2$ and $v_3$. As expected, $\hat{\rho}_{xx}(v_1,v_1,\tau)$ revealed a high degree of periodicity, which could be approximated by a regular train of impulses of period 1 s. This observation agrees with the simulated stimulation rate of 1 Hz. By contrast, $\hat{\rho}_{yy}(v_1,v_1,\tau)$ showed a lower degree of periodicity since $J_y(v,t)$ components were activated in a more sporadical manner.
The normalized cross-correlations $\hat{\rho}_{xx}(v_1,v_2,\tau)$ and $\hat{\rho}_{xx}(v_1,v_3,\tau)$ also consisted of a regular train of impulses of period 1 s and additionally, their maximum values occured at  $\tau=0.004$ s and $\tau=0.048$ s respectively. These delays corresponded to the traveling times of the excitation waves from $v_1$ to $v_2$ and from $v_1$ to $v_3$, respectively. Thus, $\hat{\rho}_{xx}(v_1,v_2,\tau)\simeq \hat{\rho}_{xx}(v_1,v_1,\tau-0.004)\simeq \hat{\rho}_{xx}(\tau-0.004)$ and $\hat{\rho}_{xx}(v_1,v_3,\tau)\simeq \hat{\rho}_{xx}(v_1,v_1,\tau-0.048)\simeq \hat{\rho}_{xx}(\tau-0.048)$. In general, the maximum value of the cross-correlation between any two dipoles occurred at a time delay $\zeta$ that corresponded to the traveling time of the excitation wave from one dipole to the other, and also $\hat{\rho}_{xx}(v,w,\tau)\simeq \hat{\rho}_{xx}(v,v,\tau-\zeta)=\hat{\rho}_{xx}(\tau-\zeta)$. 
This observation confirmed that this simulated dynamics followed a FC model [c.f. \eqref{eq:autocorrelation_correlated1}].
Finally, we observed that $\hat{R}_{J}(\tau)$ also exhibited a high degree of periodicity (Fig. \ref{fig:Plano_rho} (e)), which was mostly due to the highly periodic activity of
$\hat{\rho}_{xx}(v_1,v_1,\tau)$. Accordingly, $S_{J}(f)$ consisted of a sequence of harmonic frequencies separated by 1 Hz (Fig. \ref{fig:Plano_histo} (c)).

\subsubsection{Fully uncorrelated spatiotemporal dynamics}
In this dynamics, all the dipoles shared approximately the same normalized autocorrelation and total autocorrelation, $\boldsymbol{\hat{\rho}}(v,v,\tau)\simeq\boldsymbol{\hat{\rho}}(\tau)$ and $\hat{R}_{J}(v,v,\tau)\simeq\hat{R}_{J}(\tau)$. Thus, this dynamics could be classified as ID [c.f. \eqref{corr_IDS} and \eqref{total_corr_IDS}]. Furthermore, $\boldsymbol{\hat{\rho}}(\tau)$ and $\hat{R}_{J}(\tau)$ were found to lack any periodicity and consisted of a single impulse located at $\tau=0$, as illustrated in Fig. \ref{fig:Plano_rho} (d) and (f). As opposed to the FC dynamics, $P_x(v)$ and $P_y(v)$ were approximately the same. This observation is in agreement with the fragmented and random nature of this simulated dynamics, in which dipoles were oriented with equal probability in every direction.  

The cross-correlation between dipoles decreased rapidly with the distance that separated them. Figure \ref{fig:Plano_rho} (d) shows that $\hat{\rho}_{yy}(v_1,v_2,\tau)$ consisted of a small but distinct spike, whereas $\hat{\rho}_{xx}(v_1,v_3,\tau)$ and $\hat{\rho}_{yy}(v_1,v_3,\tau)$ were practically zero. In other words, this dynamics exhibited highly localized spatiotemporal correlations. Accordingly, although this dynamics was approximately FU, it could be more accurately described as \emph{partially correlated} (PC). We estimated the region whose dipoles showed a high degree of correlation with the dipole at $v_1$, and found it corresponded approximately to $A_1$. Interestingly, ${\hat{\rho}}_{xx}(v_1,v_2,\tau)\simeq 0$, and the maximum of ${\hat{\rho}}_{yy}(v_1,v_2,\tau)$ occurred at $\tau=0$, as seen in Fig. \ref{fig:Plano_rho} (d). In words, only the $y$ components of dipoles at $v_1$ and $v_2$ were correlated and they showed frequent simultaneous activities. The explanation for this observation is the following. Dipoles at $v_1$ and $v_2$ shared the same $x$ coordinate and they were simultaneously activated whenever local excitation wavefronts traveled along the $y$ direction, which favored the generation of $y$ dipole components. As a consequence, the $y$ components of cardiac dipoles at $v_1$ and $v_2$ were often activated simultaneously and hence, ${\hat{\rho}}_{yy}(v_1,v_2,\tau)$ showed its maximum at $\tau=0$.

\begin{figure*} [t]
\centering
\begin{tabular}{cc}
\epsfig{file = 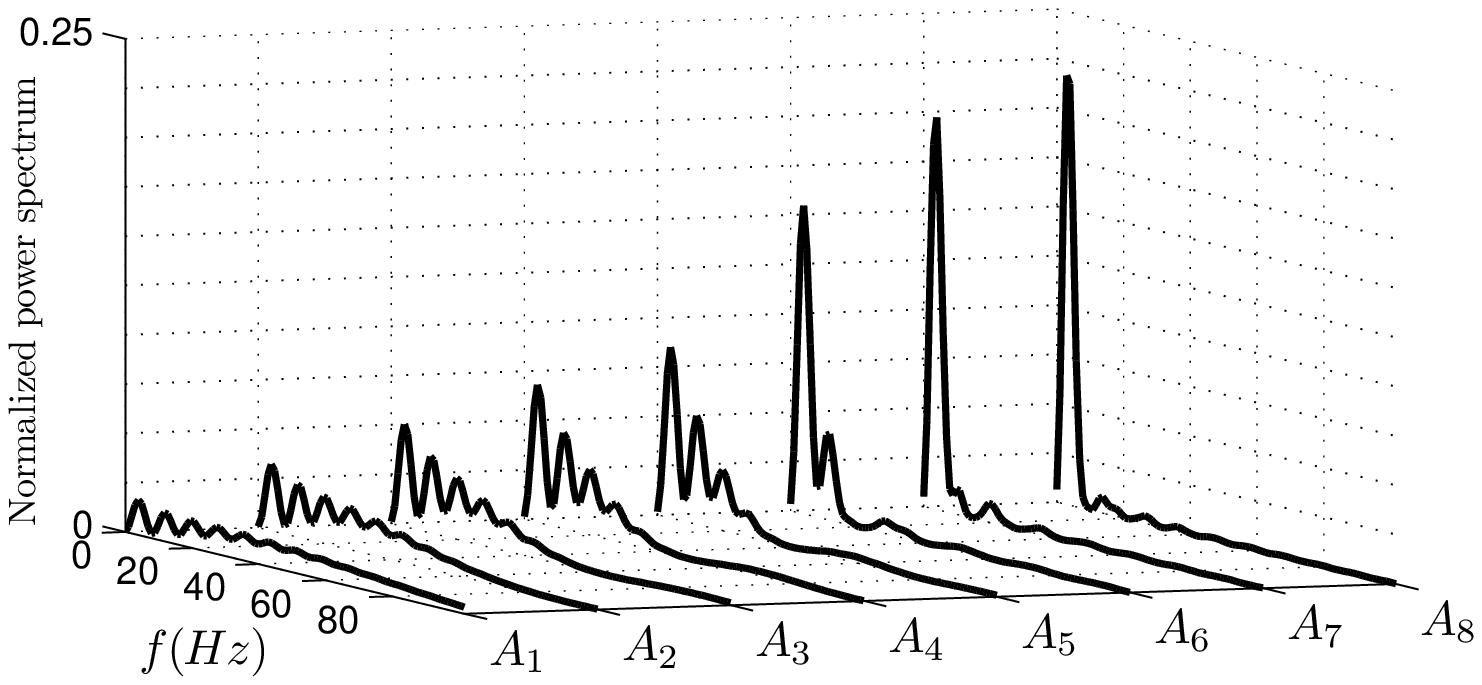,width = 10cm} \\
(a)\\
\epsfig{file = 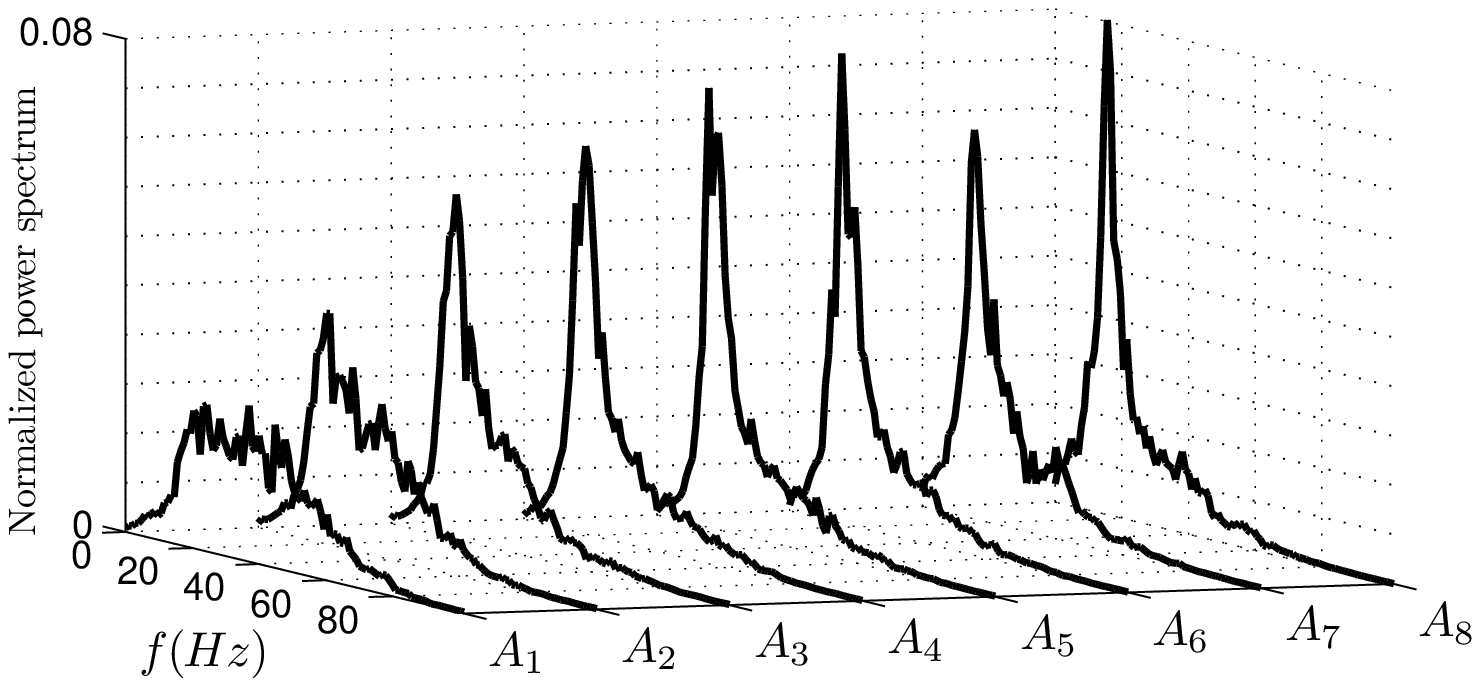,width = 10cm} \\
(b)\\\\
 \end{tabular}
 \caption{The spectral effects of the characteristics of pulse MSD are qualitatively different during FC and FU dynamics. (a) During FC dynamics, the smaller the pulse, the broader the bandwidth of synthesized leads. (b) During FU dynamics, spectra remain approximately unchanged across different leads except the smallest ones, $A_1$ and $A_2$. This behavior can be explained by the existence of localized spatiotemporal correlations in the simulated FU dynamics.}
  \label{fig:Plano_pwelch}
\end{figure*}

\subsection{Analysis of the TDDF during FC dynamics}\label{sec:TDDF}

Following \eqref{spectrum_FC_CS_2}, the spectrum of cardiac signals measured by pulse MSD during FC dynamics can be expressed as the product of the total spectrum by the Fourier Transform of the TDDF. We used our simulations to compare the spectrum of synthesized cardiac signals during FC dynamics, $S_{c_n}(f)$, and the spectrum predicted by \eqref{spectrum_FC_CS_2}. Firstly, we estimated spectra $S_{c_n}(f)$ by applying to $c_n(t)$ Welch's method \cite{Proakis06}, with a 2048-samples Hamming window and 50\% overlap. This configuration resulted in a high spectral resolution suitable to identifying the harmonic structure of $c_n(t)$. We subsequently built the time-delay function ${\zeta}_{A_n}(v,w)$ for each $A_n$, by computing the difference between the excitation times of each pair of cells within $A_n$. Then, we obtained the TDDF $F_{A_n}(\zeta)$ from ${\zeta}_{A_n}(v,w)$, and we calculated its Fourier Transform, $S_{A_n}(f)$. Finally, we obtained the spectrum predicted by \eqref{spectrum_FC_CS_2} by estimating the total spectrum of a single dipole $S_{J}(f)$ and multiplying it by $S_{A_n}(f)$. In all cases, the spectrum obtained by virtue of \eqref{spectrum_FC_CS_2} matched the spectrum of synthesized cardiac signals.

We illustrate these results with pulse MSD $A_5$ and $A_8$. Figure \ref{fig:Plano_histo} (a) shows, on an isochronal map of cell excitations, the $21\times21$ pulse corresponding to $A_5$, and the $81\times81$ pulse corresponding to $A_8$. The excitation times of four isochrones are also indicated, namely $t=0.01,0.05,0.07$ and 0.11 s. According to this map, the earliest excitation in pulse $A_5$ ($A_8$) was $t_e=0.05$ s ($t_e=0.01$ s), whereas the latest excitation was $t_l=0.07$ s ($t_l=0.11$ s). Thus, the maximum time delay between excitations in pulse $A_5$ ($A_8$) was $\zeta_{max}=t_l-t_e=0.02$ s ($\zeta_{max}=t_l-t_e=0.1$ s). Also, since the time delay function satisfies  $\zeta(v,w)=-\zeta(w,v)$, the minimum time delay in $A_5$ ($A_8$) was $\zeta_{min}=-0.02$ s ($\zeta_{max}=-0.1$ s) and thus, the possible values of time delays within $A_5$ ($A_8$) lied in the range $|\zeta|\leq 0.02$ s ($|\zeta|\leq 0.1$ s).

The generated TDDF, $F_{A_5}(\zeta)$ and $F_{A_8}(\zeta)$, are shown in  Fig. \ref{fig:Plano_histo} (b). Because of the property $\zeta(v,w)=-\zeta(w,v)$, they were symmetric around zero. In addition to this, they exhibited a triangular shape and $F_{A_5}(\zeta)=0$ ($F_{A_8}(\zeta)=0$) for time delays $|\zeta|\geq 0.02$ s ($|\zeta|\geq 0.1$ s). In all cases, we found that $F_{A_n}(\zeta)$ had a triangular shape, which is a consequence of the constant speed of the excitation wave. We also observed that the smaller $A_n$, the narrower $F_{A_n}(\zeta)$, since the maximum difference of time delays was also smaller. Accordingly, $S_{A_n}(f)$ had a square-sinc shape and the smaller $A_n$ the broader $S_{A_n}(f)$. As illustrated in Fig. \ref{fig:Plano_histo} (c), $S_{J}(f)$, $S_{c_5}(f)$ and $S_{c_8}(f)$ shared the same harmonic structure, which consisted of harmonic frequencies located at multiples of 1 Hz. This is in agreement with the fundamental periodicity imposed by the simulated stimulation rate. Figure \ref{fig:Plano_histo} (c) also shows that the envelopes of $S_{c_8}(f)$ and $S_{c_5}(f)$ conformed respectively to $S_{A_8}(f)$ and $S_{A_5}(f)$.

\subsection{Analysis of the spectral effects of lead systems}\label{subsec:spectrum}

According to \eqref{spectrum_FC_CS_2} and \eqref{eq:spectrum_FUS_IDS}, the spectral effects of pulse MSD are qualitatively different for FC and ID-FU dynamics. In order to explore our analytical results, we compared the spectra obtained by pulse MSD $A_n$ during the simulated FC and FU dynamics. We estimated the spectrum of each synthesized cardiac signal based on Welch's method, using a 512-samples Hamming window and 50\% overlap. This configuration resulted in a low spectral resolution suitable to study the spectral envelope. Based on the estimated spectra, we computed the 95\% power bandwidth of synthesized signals during FC and FU dynamics (Table \ref{Tabla_Pulse}).

During FC dynamics, we observed that as the size of $A_n$ increased, the bandwidth of synthesized cardiac signals decreased accordingly (Fig. \ref{fig:Plano_pwelch} (a)). In order to assess quantitatively the effects of the LEV on the spectrum of cardiac signals, we compared the 95\% power bandwidth of synthesized signals against the LEV (Table \ref{Tabla_Pulse}). This comparison shows that during FC dynamics, the larger the LEV, the narrower the bandwidth. By contrast, during FU dynamics the spectrum remained approximately unchanged across the different leads, except for the smallest pulses, $A_1$ and $A_2$ (Fig. \ref{fig:Plano_pwelch} (b)). Accordingly, the 95\% power bandwidth remained approximately unchanged and discrepancies were observed only for the smallest LEV (Table \ref{Tabla_Pulse}). The changes in the spectrum of cardiac signals synthesized by pulse MSD with small LEV during FU dynamics could be ascribed to the previously identified, localized spatiotemporal correlations. Within small regions of the cardiac source, it could no longer be assumed that the activities of the dipoles were uncorrelated and hence, signals measured by pulse MSD whose LEV was small enough, would exhibit a spectrum dependency on the LEV, characteristic of correlated sources. In our simulations, the estimated volume of correlation during FU dynamics was close to $A_1$ and so, the effects of local correlations manifested more clearly in $A_1$ and $A_2$.

\section{\label{sec:discussion} Conclusions and discussion}

In this paper we have presented a mathematical formalism for investigating the spectrum of cardiac signals. Our formalism reveals that the spectrum of cardiac signals can be expressed in terms of the spectrum of the underlying cardiac source and the MSD of the lead system [c.f. \eqref{eq:signal_model_spectrum}]. Two main conclusions can be drawn from \eqref{eq:signal_model_spectrum}. Firstly, the information contained in the spectrum of cardiac signals is limited to the second-order characteristics of cardiac sources. And secondly, for the same underlying spatiotemporal dynamics, different lead systems will, in general, produce cardiac signals with different spectra. Although the distortion introduced by the lead system follows in a natural way from the linearity of the measurement \eqref{eq:EqSintesis}, to the best of our knowledge this has not been formulated before. 

To gain understanding of the spectrum of cardiac signals, we have further analyzed \eqref{eq:signal_model_spectrum} for specific models of MSD and cardiac sources. Our analytical results show that the spectral effects of pulse lead systems is qualitatively different during FC and FU dynamics. During FC dynamics the spectrum of cardiac signals depends on the lead SR, which we quantified with the LEV [c.f. \eqref{spectrum_FC_CS_2}]. Specifically, the spectral envelope of cardiac signals from local measurements (small LEV) is broad, whereas the spectral envelope from global measurements (large LEV) is narrow. By contrast, during FU dynamics the spectrum of cardiac signals is a volume average [c.f. \eqref{eq:spectrum_FUS}] and hence, during ID-FU dynamics it remains the same irrespective of the LEV [c.f. \eqref{eq:spectrum_FUS_IDS}]. We have further explored the spectral manifestation of FC and FU dynamics in a simulation environment and our simulation results are in agreement with the analytical ones. 

Based on our simulation experiments, we could also explore the spectral manifestation of PC sources. In the analysis of the spatiotemporal characteristics of simulated FU dynamics, we observed localized spatiotemporal correlations which allowed us to more accurately describe the simulated dynamics as PC. Also, when analyzing the spectrum of cardiac signals synthesized from this dynamics, we observed spectral distortions for small LEV. These observations could be explained as follows. For large LEV, lead systems measure uncorrelated dipoles, and no spectral distortions are observed. However, for small LEV lead systems focus within regions where the dipoles are in fact correlated and accordingly, some distortions on the spectrum of cardiac signals are expected. Our observations encourage us to speculate that in general, irrespective of the type of MSD, the lead SR is a major factor in determining the spectrum of cardiac signals. We have derived analytically this result for idealized pulse MSD during FC and FU dynamics, but not for generic MSD nor for PC dynamics. We can speculate that in those scenarios where the LEV is larger than the volume of correlation, PC dynamics can be treated as FU, whereas whenever the LEV is within the volume of correlation, PC dynamics can be treated as FC. Nevertheless it remains to further study the spectrum of cardiac signals for generic lead systems and generic spatiotemporal dynamics. The formalism that we have presented constitutes a convenient mathematical framework for investigating the spectrum of cardiac signals measured during generic spatiotemporal dynamics. Within our formalism, a meaningful definition of the degree of spatiotemporal correlation and the volume of correlation could be proposed, and by using \eqref{eq:signal_model_spectrum} its effects on the spectrum of cardiac signals could be thoroughly investigated.

The study that we have presented has practical implications on the analysis of the spectrum of cardiac signals. We have shown that during correlated dynamics leads can distort the spectral envelope, whereas the harmonic separation remains unchanged. Therefore, spectral features which depend on the spectral envelope such as the peak frequency \cite{Eftestol00,Jekova04}, the mean frequency \cite{Dzwonczyk90}, and the organization index \cite{Everett01a}, can be affected by the characteristics of the lead system, whereas those features that only depend on the harmonic separation, notably the DF \cite{Skanes98,Mandapati00,Mansour01,Sanders05,Atienza06}, are insensitive to the characteristics of the lead system. This conclusion supports, from a theoretical standpoint, recent experimental studies which assessed the robustness of the DF. In \cite{Berenfeld11} it was reported that the values of the DF obtained from unipolar and bipolar cardiac signals correlate highly with the DF values obtained from optical signals. Other studies have compared the DF values obtained from cardiac signals measured by contact and non-contact intracardiac mapping systems, and have found good agreement between the DF values obtained from each mapping system \cite{Lin07,Gojraty09}. Finally, a high correlation between DF values from intracardiac EGM and surface ECG signals was found in \cite{Dibs08}. In summary, these experimental studies indicate in agreement with our theoretical results, that the DF is a robust spectral feature of cardiac signals. Therefore, there is both experimental and theoretical evidence of the robustness of the DF, which pave the ground for developing novel DF mapping systems for diagnosing and managing of fibrillation, such as non-contact, intracardiac DF mapping systems \cite{Lin07,Gojraty09} and non-invasive, surface DF mapping systems \cite{Berenfeld10}.

\let\thefigureSAVED\thefigure 
\let\thetableSAVED\thetable 

\appendix
\section{\label{sec:appendixA} Derivation of the Correlation and the Spectrum of Fully Correlated Sources}

By setting $v=w$ in \eqref{eq:autocorrelation_source} and substituting \eqref{def:FCS} into it, it can be proved that the autocorrelations of any two dipoles $\mathbf{J}(v,t)$ and $\mathbf{J}(w,t)$ in a FC source, are always identical. 
\begin{eqnarray}
\boldsymbol{\rho}(w,w,\tau)&=&\langle\mathbf{J'}^{T}(w,t+\tau)\mathbf{J'}(w,t)\rangle_t  \nonumber \\
					      &=&\langle\mathbf{J'}^{T}(v,t+\tau-\zeta (v,w))\mathbf{J'}(v,t-\zeta (v,w))\rangle_t  \nonumber \\
					     &=&\langle\mathbf{J'}^{T}(v,t'+\tau)\mathbf{J'}(v,t')\rangle_t  \nonumber \\
					      &=&\boldsymbol{\rho}(v,v,\tau)
\end{eqnarray}
where $t'=t-\zeta (v,w)$. Therefore, FC sources are by construction also ID, $\boldsymbol{\rho}(v,v,\tau)=\boldsymbol{\rho}(\tau)$ and $\boldsymbol{\sigma}(v,v,f)=\boldsymbol{\sigma}(f)$.
By substituting \eqref{def:FCS} into \eqref{eq:autocorrelation_source}, $\boldsymbol{\rho}(v,w,\tau)$ can be expressed in terms of $\boldsymbol{\rho}(\tau)$:
\begin{eqnarray}\label{eq:autocorrelation_correlated}
\boldsymbol{\rho}(v,w,\tau) &=&\langle\mathbf{J'}^{T}(v,t+\tau)\mathbf{J'}(w,t)\rangle_t  \nonumber \\
		 &=&\langle\mathbf{J'}^{T}(v,t+\tau)\mathbf{J'}(v,t-\zeta (v,w))\rangle_t   \nonumber \\
		 &=&\langle\mathbf{J'}^{T}(v,t+\tau)\mathbf{J'}(v,t)\rangle_t \ast \delta (\tau - \zeta (v,w)) \nonumber \\
                    &=&\boldsymbol{\rho}(\tau) \ast \delta (\tau - \zeta (v,w)) \nonumber \\
                    &=&\boldsymbol{\rho}(\tau - \zeta (v,w)).
\end{eqnarray}
In words, $\boldsymbol{\rho}(v,w,\tau)$ is a $\zeta(v,w)$-delayed version of $\boldsymbol{\rho}(\tau)$. As a consequence, by applying the Fourier Transform to \eqref{eq:autocorrelation_correlated}, the spectrum of a FC source can be expressed as:
\begin{eqnarray}
\boldsymbol{\sigma}(v,w,f) &=& \mathcal{F} [\boldsymbol{\rho}(v,w,\tau)] \nonumber \\
&=& \mathcal{F}[\boldsymbol{\rho}(\tau) \ast \delta (\tau- \zeta (v,w))] \nonumber \\
&=& \mathcal{F}[\boldsymbol{\rho}(\tau)] \mathcal{F}[\delta (\tau- \zeta (v,w))] \nonumber \\
&=& \boldsymbol{\sigma}(f) \exp[-j2\pi f \zeta (v,w)].
\end{eqnarray}

\section{\label{sec:appendixB} Connecting the Spectrum of Cardiac Signals to the Spectrum of Cardiac Sources}

By substituting \eqref{eq:EqSintesis} into \eqref{c_minus_average}, we can express $c'(t)$ in terms of $\mathbf{J'}(v,t)$,
\begin{eqnarray}\label{c_sin_media2}
c'(t) 		 &=& \int_V{\mathbf{L}^{T}(v) \mathbf{J}(v,t)}{dv}-\langle \int_V{\mathbf{L}^{T}(v) \mathbf{J}(v,t)}{dv} \rangle_t \nonumber \\
		&=& \int_V{\mathbf{L}^{T}(v) \mathbf{J}(v,t)}{dv}-\int_V{\mathbf{L}^{T}(v) \bar{\mathbf{J}}^{T}(v)}{dv} \nonumber \\
		 &=& \int_V{\mathbf{L}^{T}(v) (\mathbf{J}(v,t)- \bar{\mathbf{J}}(v))}{dv} \nonumber \\
		 &=& \int_V{\mathbf{L}^{T}(v) \mathbf{J'}(v,t)}{dv}. 
\end{eqnarray}
Then, by substituting  \eqref{c_sin_media2} into \eqref{signal_autocorr}, we can write the autocorrelation of a cardiac signal $R_{c}(\tau)$ in terms of the autocorrelation of the cardiac source, $\boldsymbol{\rho}(v,w,\tau)$:
\begin{eqnarray}\label{eq:signal_model_autocorr2}
R_{c}(\tau) 	&=& \langle \int_V{\mathbf{L}^{T}(v) \mathbf{J'}(v,t+\tau)}{dv} \int_V{\mathbf{L}^{T}(w) \mathbf{J'}(w,t)}{dw}\rangle_t \nonumber \\
        			&=& \int_{V\times V}{ \mathbf{L}^{T}(v) \langle \mathbf{J'}(v,t+\tau)\mathbf{J'}^{T}(w,t)\rangle_t \mathbf{L}(w) }{dvdw}\nonumber \\
        			&=& \int_{V\times V}{\mathbf{L}^{T}(v) \boldsymbol{\rho}(v,w,\tau) \mathbf{L}(w)}{dvdw}.
\end{eqnarray}
Finally, following \eqref{signal_spectrum}, we apply the Fourier Transform to \eqref{eq:signal_model_autocorr2} and derive the relationship between the power spectrum of a cardiac signal $S_{c}(f)$ and the spectrum of the cardiac source, $\boldsymbol{\sigma}(v,w,f)$:
\begin{eqnarray}
S_{c}(f) &=&\mathcal{F}\left\{\int_{V\times V}{\mathbf{L}^{T}(v) \boldsymbol{\rho}(v,w,\tau) \mathbf{L}(w)}{dv dw}\right\} \nonumber \\
        &=&\int_{V\times V}{\mathbf{L}^{T}(v) \mathcal{F}\{\boldsymbol{\rho}(v,w,\tau)\} \mathbf{L}(w)}{dv dw} \nonumber \\
        &=&\int_{V\times V}{\mathbf{L}^{T}(v) \boldsymbol{\sigma}(v,w,f) \mathbf{L}(w)}{dv dw}.
\end{eqnarray}

\let\thefigure\thefigureSAVED 
\let\thetable\thetableSAVED

\end{document}